    \documentclass[twocolumn,amsmath,amssymb,superscriptaddress]{revtex4}
\usepackage[usenames,dvipsnames]{color}
\usepackage{graphicx}
\usepackage{dcolumn}
\usepackage{bm}
\usepackage{textcomp}

\def\mclx{}
\def\mcl{}
\def\mcln{}

\def\rice{Department of Physics and Astronomy, MS 108, Rice University, Houston, Texas 77005, USA}
\def\llnl{Lawrence Livermore National Laboratory, Livermore, California 94551, USA}

\def\nc{n_{c}}
\def\up{u_p}

\def\er{\mathcal{E}}
\def\ls{\ell_{s}}
\def\ni{n_{i}}
\def\ne{n_{e}}

\def\refl{\mathcal{R}}
\def\fah{f_{h}}
\def\fap{f_{p}}
\def\gamh{\gamma_h}

\def\bp{\beta_p}
\def\bpz{\beta_{0}}
\def\dph{\Delta \phi}

\def\falim{f_{*}}
\def\rh{\overline{\rho}_h}
\def\isub{I_{18}}
\def\fs{g_s}
\def\fl{F_L}
\def\lc{\ell_{c}}
\def\fmax{f^*}
\def\favg{\langle f\rangle}

\begin{document}

\title{Kinematic Constraints on Absorption of Ultraintense Laser Light}

\author{M. C. Levy}
\affiliation{\rice}
\affiliation{\llnl}
\email{levy11@llnl.gov}
\author{S. C. Wilks}
\affiliation{\llnl}
\author{M. Tabak}
\affiliation{\llnl}
\author{S. B. Libby}
\affiliation{\llnl}
\author{M. G. Baring}
\affiliation{\rice}
\email{levy11@llnl.gov}

\date{\today}

\begin{abstract}
\mclx{We derive upper and lower bounds on the absorption of ultraintense laser light by solids as a function of fundamental laser and plasma parameters.  These limits emerge naturally from constrained optimization techniques applied to a generalization of the laser-solid interaction as a strongly-driven, relativistic, two degree of freedom Maxwell-Vlasov system. We demonstrate that the extrema and the phase-space-averaged absorption must always increase with intensity, and increase most rapidly when $10^{18} < I_L \   \lambda_L^2 < 10^{20}$ W $\mu$m$^2/$cm$^{2}$.  Our results indicate that the fundamental empirical trend towards increasing fractional absorption with irradiance therefore reflects the underlying phase space constraints.}
\end{abstract}

\maketitle

The interaction of an ultraintense laser with solid matter is characterized by the nonlinear action of the light\cite{Wilks1992}. \mclx{
A fundamental empirical property of the interaction is the trend towards increasing \mcl{fractional} absorption (percent of laser energy absorbed) with irradiance. While this has been established through simulation and experimental data over the years\cite{Ping2008}, the theoretical understanding to date has remained heuristic.
}
\mcln{As the laser strikes the target,} the ponderomotive force \mcl{$\fl$ generally} couples the incident photon flux into two kinetic modes:  'hole boring' or 'hole punching' (hp) ions accelerated by the space-charge force associated with electrons under the excursion of time-averaged field energy gradients; and relativistic 'hot' electrons excited by the oscillatory component of $\fl$ at $2 \omega_L$\cite{Wilks1992,Denavit1992}.
\mcln{The interplay between these absorption mechanisms underpins all ultraintense laser-solid applications, from relativistic particle acceleration\cite{Malka2008}, anti-matter generation\cite{Chen2010}, to fast ignition inertial confinement fusion\cite{Tabak1994}.} 

\mcl{
\mcln{Recently, we developed an analytic framework treating both ponderomotive absorption modes\cite{Levy2013PoP}, in effect extending the relativistic hp model\cite{Naumova2009,Macchi2013a} with the degree of freedom associated with electrons coupling into the high-frequency mode.
The phase space properties of such systems are of general interest, particularly the total fractional light absorbed by the plasma $f$ and the coupling efficiencies into hot electrons $\fah$ and into hp ions $\fap$.}  
The general consideration of optimal couplings under the constraint of phase space conservation further motivates these studies\cite{Fisch1993}.
}

\mclx{In this Letter, we examine the phase space constraints on the generalized ultraintense laser-solid interaction, modeled as a strongly-driven, relativistic, two degree of freedom Maxwell-Vlasov system subject to conservation laws at the laser-matter interface. One degree of freedom corresponds to the highly-relativistic hot electron component and the other to the moderately relativistic hp ion component.
We derive upper and lower bounds on the absorption of ultraintense laser light by solids as a function of fundamental laser and plasma parameters. 
These limits emerge naturally from constrained optimization techniques\cite{Mehrotra1992} applied to the system, and generally establish bounds on existing mechanisms of collisionless absorption (e.g., those reviewed in \cite{Wilks1997}).
We demonstrate that at the extrema and at the phase-space-average absorption must always increase with intensity $I_L$.  We also show that absorption must increase most rapidly when $10^{18} < I_L \   \lambda_L^2 < 10^{20}$ W $\mu$m$^2/$cm$^{2}$, the regime in which  experiments show absorption going from near-zero to near-unity\cite{Ping2008}.  Our results therefore form a plausible theoretical basis for the fundamental empirical trend towards increasing absorption with irradiance.
}

\mcl{First, we briefly review key aspects of the fully-relativistic absorption model\cite{Levy2013PoP} required for our analysis. 
In a manner analogous to the magnetohydrodynamic shock relations\cite{DeHoffmann1950}, a connection is established across the LP interface between the laser and unperturbed plasma and the properties of the \mcln{excited particle beams.  }
We express this connection using two tensor equations: one for particle number conservation; the other for the four-divergence of the electromagnetic stress-energy tensor $T^{\mu \nu}$ and a source term for particle flux,}
\begin{eqnarray}
\frac{\partial \Gamma_s^\mu}{\partial x^\mu} = 0, \quad \quad \Gamma_s^\mu = \frac{n_s'}{m_s} P_s^\mu \nonumber \\
n_s = \int \fs\ \mathrm{d} p^k , \quad P_s^\mu = \frac{1}{n_s} \int p^\mu \ \fs\ \mathrm{d} p^k \label{eqn:numConsvn} \\
\frac{\partial T^{\mu \nu}}{\partial x^\nu}+ \eta^{\mu k} \frac{\mathrm{d} p_k}{\mathrm{d} \tau} = 0 \label{eqn:stressE}
\end{eqnarray}

\mcl{Here $n_s'$ is the proper density of the particle population of type $s$ excited by the laser, $P_s^\mu = \gamma_s m_s c\ ( 1, \mathbf{V_s}/c )$ is the ensemble-average four-momentum, and the Lorentz factor $\gamma_s = (1-\mathbf{V_s}\cdot \mathbf{V_s}/c^2)^{-1/2}$.  
The term $d p^k/d \tau = -(q_s/m_s) F^{k \mu} p_\mu$ includes spatial components due to the mass-shell restriction\cite{Ruhl1995,Hazeltine2002}. $\fs(x^k, p^k)$ is the particle distribution function, $\tau$ is the proper time, $q_s$ is the electric charge, $m_s$ is the rest mass and $F^{\mu \nu}$ represents the field strength tensors.  
We adopt the the Minkowski tensor conventions $\eta^{\mu \nu}$ has signature $(-,+,+,+)$; the greek sub- and super-scripts represent tensor indices $\mu \in \{0,1,2,3\}$; and the latin indices $k \in \{1,2,3\}$ run over the spatial subset. }

\begin{figure}[t]
\begin{center}
\resizebox{8.5cm}{!}{\includegraphics{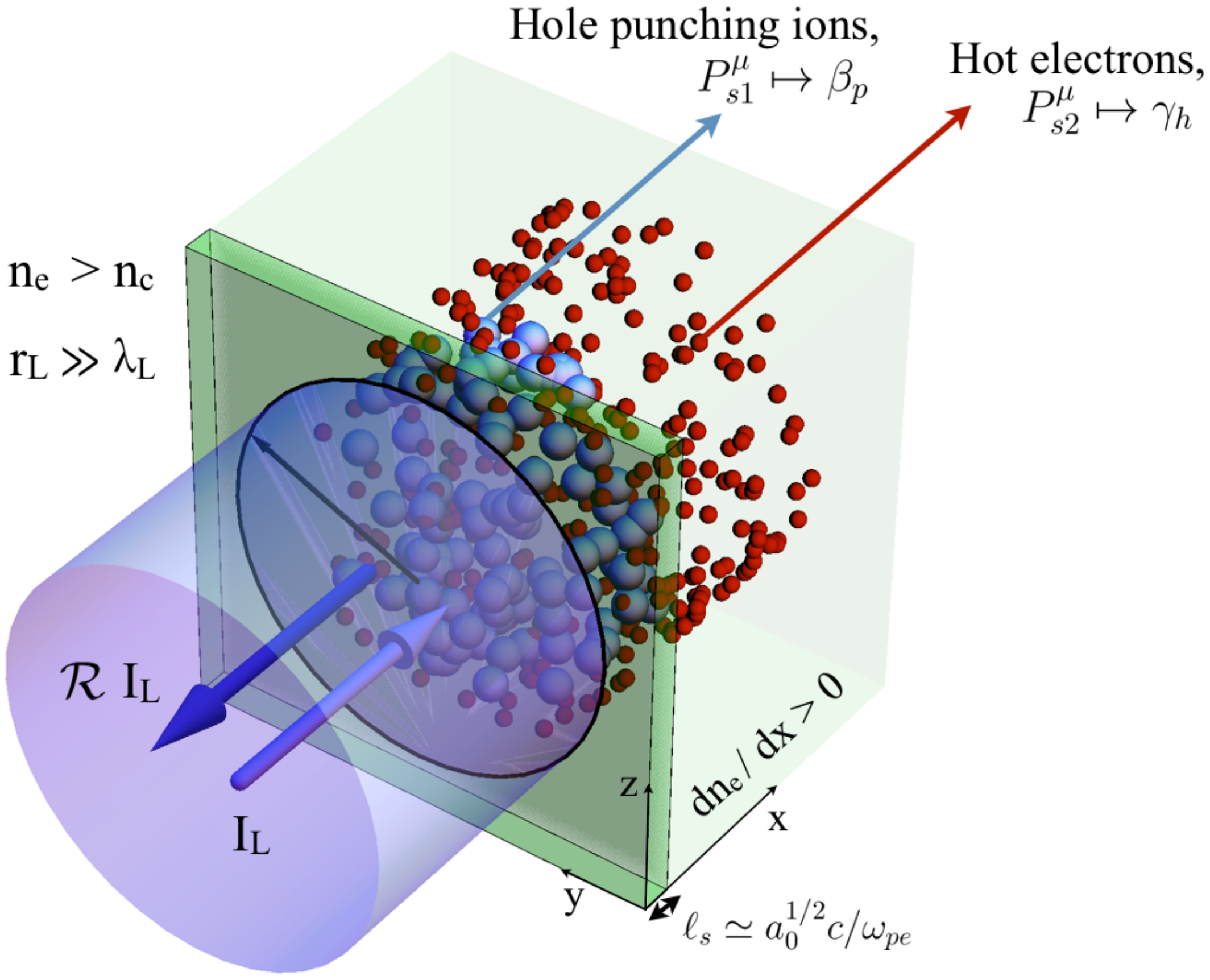}}
  \end{center}
  \caption{\mclx{Schematic of the ultraintense laser-solid interaction focusing on relativistic particle excitation in the Lorentz-transformed collisionless skin depth $x\in[0,\ls]$.}} 
  \label{fig:schem} 
\end{figure}

\mcl{Given this description of the ponderomotive coupling, there are two  exchange-mediating populations, i.e., those excited by the laser on the spatial scale $\ls \simeq a_0^{1/2} c/\omega_{pe}$ at the overdense\cite{Cattani2000} LP interface where $a_0 = e E_L/(m_e c \omega_L)$ is the normalized laser vector potential ($e$ is the electric charge, $E_L$ is the laser electric field and $\omega_L$ is the laser frequency). 
They are enumerated by the set
$s = \{ \text{hp ions, hot electrons} \}$. For simplicity we neglect the hp electrons (associated with the space-charge force accelerating ions) in $s$ as their calculational role is suppressed by the electron-ion mass ratio.
This allows us to focus on two degrees of freedom. In the following these species will be referred to by their indices.
We assume that the hot electron component is comprised of particles coupling into the high-frequency mode of the nonlinear force, with ensemble-average velocity $V_{s2} \simeq c$, though the analysis presented here works for general acceleration microphysics (\cite{Haines2009}, in analogy to \cite{DeHoffmann1950}).
}

\mcl{Using equations (\ref{eqn:numConsvn}-\ref{eqn:stressE}), we examine the steady-state quasi-1-D laser-solid interaction.  This case is significant as it bounds the true three-dimensional situation.  Detailed discussion of this point follows at the end of this Letter.  Physically, an ultraintense laser having infinite transverse extent, intensity $I_L$ and wavevector $\mathbf{k_L}$ is normally incident onto an overdense target with arbitrary, slowly-changing heterogenous plasma density profile\cite{Naumova2009}.    All particles in the LP interface are assumed to interact with the laser and accelerate by coupling into the modes associated with set $s$. The 'laser piston' sweeps up and fully reflects hp particles along $\hat{k}_L$\cite{Naumova2009}, and the hot electron beam is assumed to be axisymmetric with ensemble-average velocity $\mathbf{V_{s2}} \cdot \hat{k}_L \approx c$.}
\mclx{Key aspects of the interaction are depicted schematically in Fig. \ref{fig:schem}.}
\mcl{Due to the fact that the momenta $P_s^\mu$ must simultaneously satisfy equation (\ref{eqn:stressE}), it is clear that only specific combinations of beam properties can self-consistently conserve energy and momentum with the laser.  The allowable hot electron and hp ion beams, in terms of the laser and unperturbed plasma quantities, thus form the basis for our phase space analysis. Employing the reduced, scaled forms $P_{s1}^\mu \mapsto \bp$ where $\bp=\up/c$ is the piston velocity, and $P_{s2}^\mu \mapsto \gamh$ where $\gamh$ is the ensemble-average hot electron Lorentz factor, the solutions can be written\cite{Levy2013PoP},}
\begin{eqnarray}
\gamh & = & \frac{(1-\refl)
   \sqrt{\bpz^2 \refl+1} - \bpz \sqrt{\refl} (1+\refl)}{\sqrt{\bpz^2
   \refl+1}\ \bpz^{-2}}\ \rh^{\ -1} + O(1) \nonumber \\
\bp & = & \bpz \ \left( \frac{\refl}{1 + \refl \bpz^2} \right)^{1/2}
\label{eqn:bp}
\label{eqn:gamH}
\end{eqnarray}
\mcl{where $\refl=1-f$ is the fractional light reflected from the LP interface, and $\bpz \equiv \left[Z m_e \nc/(2 M_i \ne)\right]^{1/2} a_0$ is the dimensionless shock velocity scale. The electron density in the LP interface is $\ne$, the ion density $\ni=\ne/Z$ assuming uniform charge state $Z$, $m_e$ is the electron mass, $M_i$ is the ion mass and $\rho_s = n_s m_s$ is the mass density. $\rh \equiv \rho_h/\sum_s \rho_s \ll 1$ is a small parameter corresponding to the relative mass density of hot electrons in the LP interface.
 The $O\left(1 \right)$ term in equation (\ref{eqn:gamH}) associated with a series expansion in $\rh$ is a polynomial in $\bpz$ and $\refl$.
Conservation of particle number in equation (\ref{eqn:numConsvn}) implies a separation of velocity scales between hp ions and hot electrons, the latter being highly-relativistic for $a_0>1$. The parameter $\rh$ reflects the importance of these relativistic effects in the phase space analysis. 
}

Conversion efficiencies can be calculated from equations (\ref{eqn:stressE}-\ref{eqn:gamH}) as,
\begin{eqnarray}
\fah & = & \frac{(1-\refl) \sqrt{\bpz^2 \refl+1} -  (1+\refl) \bpz \sqrt{\refl}}{\sqrt{\bpz^2
   \refl+1}-\bpz \sqrt{\refl}} + O\left( \frac{\rh}{\bpz^2} \right) \nonumber \\
\fap & = & \frac{2 \bpz \refl^{3/2}}{\sqrt{\bpz^2 \refl+1}-\bpz \sqrt{\refl}}
\label{eqn:fah}
\label{eqn:fap}
\end{eqnarray}
\mcl{where $\fah$ and $\fap$ are respectively the hot electrons and hp ion absorption fractions.
Equation (\ref{eqn:fap}) generally predicts finite total absorption $f>0$ in the $M_i \rightarrow \infty$ limit. 
Equation (\ref{eqn:fap}) also highlights the key result that the phase space variables $(\bpz,\refl)$ control all ultraintense-solid interactions for $\rh/\bpz^2 \ll 1$.  As a corollary, electron dynamical effects, due to light polarization, enter the equations indirectly through the phase parameter $\refl$. }

\mcl{Having obtained the conversion efficiencies for both the steady-state and oscillatory coupling, we recast the LP interaction in a novel variational framework. First consider a circularly-polarized laser pulse interacting with an overdense target. In this situation, the total absorbed light
 $f=\fap$ \cite{Naumova2009}.  
 Now, by 'detuning' the relative phase between the linearly-polarized waves comprising the circular light away from $\pi/2$, the ponderomotive force obeys  $|\langle\mathbf{\fl}\rangle-\mathbf{\fl}| > 0$ and light may couple into oscillatory mode hot electrons.  
Because $\fah$ and $\fap$ are generally $> 0$, their nonlinear, constrained relationship can be explored to yield a global absorption extremum.}

\begin{figure}[t]
\begin{center}
\resizebox{8cm}{!}{\includegraphics{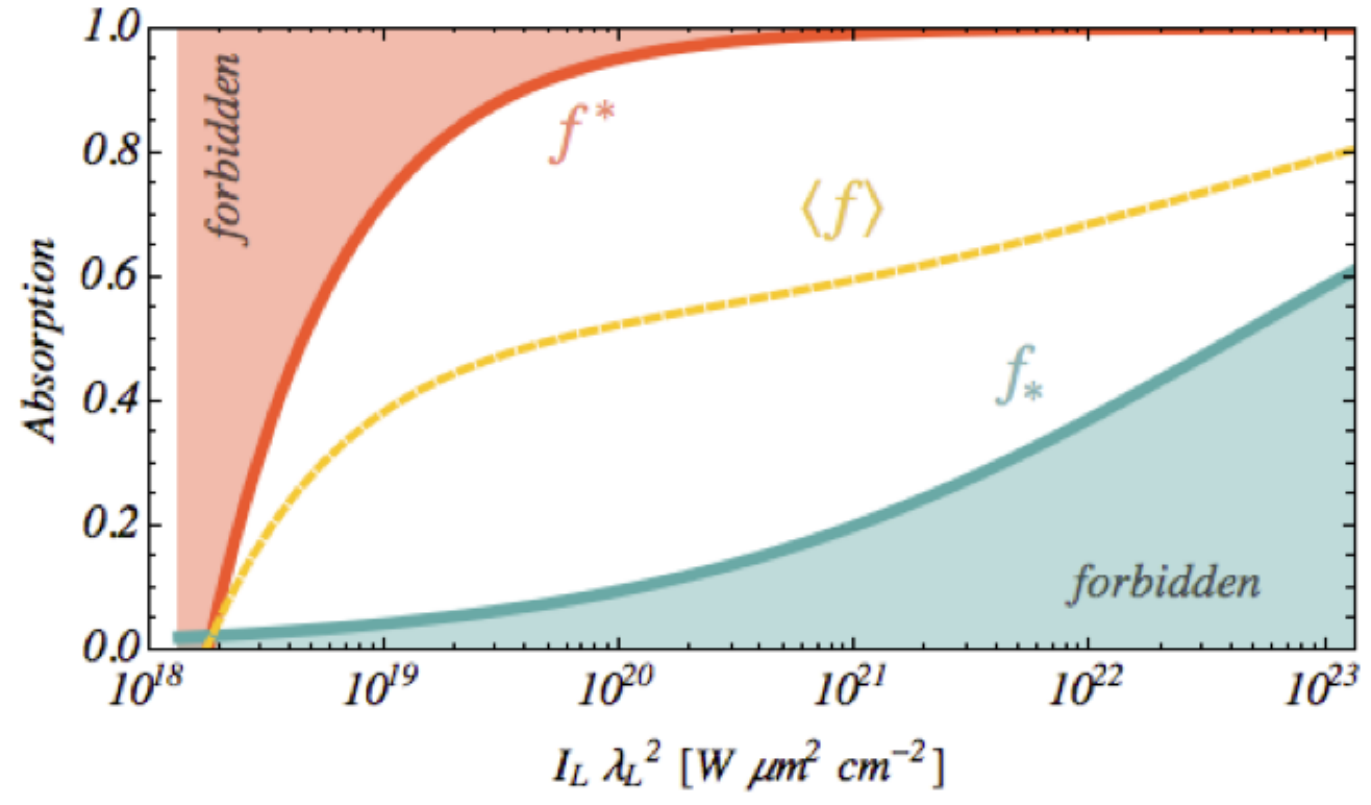}}
  \end{center}
  \caption{\mclx{Extrema and phase-space-averaged absorption.}}
  \label{fig:abs} 
\end{figure}

Since  $\gamh \ \geq 1$, by considering this constraint in detail we may derive an expression for the lower limit on light absorption. \mclx{Contours of $\gamh$ in equation (\ref{eqn:gamH}) exhibit a separatrix representing the physically allowed boundary of $\bpz$ such that $\gamh-1=0$. } In the framework of constrained minimization\cite{Mehrotra1992}, the lower limit on absorption, \mcl{$\falim$}, can be rigorously defined as,
\begin{eqnarray}
\falim (\bpz) & \equiv & \text{Min} \left( \fah + \fap \right) \nonumber \\
\text{s.t.} & & \refl \in [0,1] \nonumber \\
  & & \gamma_h\ge 1 \label{eqn:fmin}
\end{eqnarray}
\mcl{Because the objective function is generally nonlinear in the phase space variables, we solve equation (\ref{eqn:fmin}) numerically by means of cylindrical algebraic decomposition\cite{Collins1975} in order to account for regions where $\rh/\bpz^2 \sim 1$.  \mclx{For a typical value of $\rh$} and a fully-ionized $Z/A=1/2$ plasma ($A=M_i/m_p$, $m_p$ is the proton mass) relevant to the LP interface, $\falim$ can be approximated as,}
\begin{eqnarray}
\falim = 1.9 \bpz - 2.75 \bpz^2 + 1.91 \bpz^3, \quad \bpz \lesssim 0.5
\label{eqn:rMaxPoly}
\end{eqnarray}
\mclx{To gain physical insight into equation (\ref{eqn:rMaxPoly}), we consider the limit of $\rh/\bpz^2 \ll 1$.  Here $\falim$ converges with the absorption in the ion-only case\cite{Naumova2009}, exhibiting $\falim \rightarrow 2 \bpz/(1+2\bpz)$.  In the two degree of freedom model, this self-consistently emerges from the unitary Lorentz factor constraint on hot electrons.} 

\mclx{To elucidate the intensity dependence of $\falim$, we employ the overdense plasma threshold condition\cite{Cattani2000}, $a_0^2 \lesssim (27/64) \left(\ne/\nc\right)^4$ for $\ne/\nc\gg 1$, to derive an \mcln{approximate} analytic form in terms of $I_L$,}
\begin{eqnarray} 
\falim = \frac{2^{3/4} \ 3^{3/8} \ (Z m_e/M_i)^{1/2} \ \isub^{3/8} }{2 + 2^{3/4} \ 3^{3/8} \ (Z m_e/M_i)^{1/2} \ \isub^{3/8}}
\label{eqn:faLimExact}
\end{eqnarray}
where \mcln{$\isub=I_L \lambda_L^2/(1.34\times 10^{18}$ W $\mu$m$^2/$cm$^{2}$)}.   
\mcln{As the laser intensity increases for fixed target density, equations (\ref{eqn:rMaxPoly}-\ref{eqn:faLimExact}) show that an increasing fraction of light must be absorbed by the plasma as a consequence of self-consistent energy and momentum conservation. 
\mclx{
Momentum imparted to the plasma decreases with absorption, leading to an increase in $\bp$ and $\fap$ according to equations (\ref{eqn:bp}) and (\ref{eqn:fap}).  Therefore equations (\ref{eqn:rMaxPoly}-\ref{eqn:faLimExact}) can be interpreted as a constraint on the momentum flux transferred from the laser to the plasma, with absorption below $\falim$ forbidden as coupling into ions $\fap$ would exceed $f$. 
}

\mclx{Using constrained optimization techniques, a maximum in absorption $\fmax$ can also be derived.  The upper limit can be expressed exactly using our phase space variables, as $\fmax=1-\rh \ \bpz^{-2}$.  In terms of intensity, $\fmax$ can be expressed using the overdense condition as,}
\begin{eqnarray} 
\fmax = 1-\frac{2^{3/2}\ \rh}{3^{3/4}\ \isub^{3/4}\ Z m_e/M_i}
\label{eqn:faUp}
\end{eqnarray}
\mclx{A physical explanation for equation (\ref{eqn:faUp}) emerges naturally and self-consistently from $\gamh$ in equation (\ref{eqn:gamH}). The upper limit on laser light absorption  $\fmax$ is associated with the excitation of hot electrons in the LP interface to the full ponderomotive potential of the laser, $1/2 \  a_0^2 m_e c^2$. 
}

\mclx{Fig. \ref{fig:abs} depicts the extrema in ultraintense laser absorption using equations (\ref{eqn:faLimExact}-\ref{eqn:faUp}), where we have exploited the relativistic nature of the model to consider cases extending out to large $I_L$.   We choose $Z/A=1/2$ and the $\rh$ value associated with all electrons in the interface coupling into the high-frequency mode.  $\favg \equiv \int f d\refl / \int d\refl$ represents the parameter-space-averaged absorption, where $f$ is the sum over equation (\ref{eqn:fah}) and the integrals are bounded by $\falim$ and $\fmax$.  We emphasize that $\favg$ is not a prediction of absorption for a given experiment, but rather, the absorption for a given intensity-density combination averaged over all possible outcomes (i.e., in relation to laser polarization).  The lower limit on absorption is dominated by the hp ions and the upper limit is dominated by the hot electrons.  Across all intensities, the hot electron degree of freedom plays an essential role, allowing the system to access absorption states lying between the extrema.  The trends shown in Fig. \ref{fig:abs} are insensitive to choice of (overdense) target density.}

In contrast to intensities from $10^{14}$ to $10^{18}$ W/cm$^2$, a fundamental property of ultraintense laser interaction with matter is the trend towards increasing absorption with $I_L$\mcl{, e.g., as reported on experimentally for linearly-polarized light in \cite{Ping2008}.}  \mclx{This trend emerges naturally from the properties of equations (\ref{eqn:faLimExact}-\ref{eqn:faUp}),}
\begin{eqnarray} 
\frac{\partial}{\partial I_L} \left( \falim, \favg, \fmax \right)  & > & 0, \nonumber \\ 
\lim_{I_{18} \rightarrow 10^3}\ \frac{\partial^2 }{\partial I_{18}^2}  \left( \falim, \favg, \fmax \right) & = & 0
\label{eqn:dlims}
\end{eqnarray}
\mclx{Equation (\ref{eqn:dlims}) shows that at the extrema and parameter-space-average, absorption must always increase with intensity in the ultraintense regime, and increase most rapidly when $10^{18} < I_L \   \lambda_L^2 < 10^{20}$ W $\mu$m$^2/$cm$^{2}$.}

\mclx{Laser-solid interactions generally satisfy $\falim \leq f \leq \fmax$.  It is useful to express this inequality in terms of the laser intensity and unperturbed plasma density,}
\begin{eqnarray} 
   \frac{6.69\times 10^{-1}
   \sqrt{\frac{ I_L [W/cm^2] }{ n_e[ cm^{-3} ] }}}{ 
   1+6.69\times 10^{-1} \sqrt{\frac{ I_L [W/cm^2] }{n_e[ cm^{-3} ]}}}
\leq f \leq \nonumber \\ 
1- 1.22 \times 10^{-3} \frac{n_e[ cm^{-3}] }{I_L [W/cm^2]}
\label{eqn:faIneq}
\end{eqnarray}
\mclx{Equation (\ref{eqn:faIneq}) is maintained for arbitrary laser polarization.}

\mcln{In order to exhibit the dynamic 'observables' associated with equations (\ref{eqn:rMaxPoly}-\ref{eqn:faUp}), we have performed particle-in-cell (PIC) simulations using the hybrid LSP code\cite{Welch2004}. Here the simple steady-state, quasi-1-D model is relaxed and we show that key features remain true in the presence of more physically realistic conditions. One representative simulation is examined here; detailed discussion and additional PIC results can be found in \cite{Levy2013PoP}.}

\mcl{In this example, LSP is configured in 1D3V (one coordinate in space, three coordinates in velocity)} Cartesian geometry.
\mcl{Laser light enters the simulation at the $x=0$ boundary and is incident upon an overdense {$Z/A = 1$} plasma slab at $x=[5,290] \mu$m.}  
The laser pulse has $1 \mu$m wavelength and rises over 3 optical cycles to a flat-top profile with 500-700fs duration.  
Light coupling into the oscillatory and steady-state absorption modes is illustrated in Fig. \ref{fig:sims} for a $3 \times 10^{22}$ W/cm$^2$ laser beam interacting with a $\ne/\nc =50$ slab (corresponding to $\bpz=0.35$).  \mcl{The light has an offset of $\dph\ = 0.8$ rad between the electric field phases (compared to circular polarization at $\dph\ = \pi/2$), allowing electrons to couple into the high-frequency mode with three degrees of freedom in velocity space.}
(a) depicts the density of exchange mediating electrons (red) and of ions (black) at two times.  The density of electrons coupled into the oscillatory mode is calculated in the simulation as the subset of exchange-mediating electrons passing through the $x=100\mu$m plane. \mcl{This diagnostic allows discrimination between the hot electrons and the hp electrons, as the former propagate at $\simeq c$.}
(d) compares simulation oscillatory mode coupling to the analytical model in detail.  The dashed black line corresponds to equation (\ref{eqn:fah}) using the average reflection coefficient $0.57$. The hot electron energy flux density $\mathrm{d}\er_h/(\mathrm{d}A\ \mathrm{d}t)$ (red) is self-consistently calculated through a diagnostic in the simulation.   The ensemble average $\gamh$ and $\rh$ from the simulation are used to calculate the solid black curve.   As illustrated in Fig \ref{fig:sims}. (c-d), the analytic model and simulation results are in excellent agreement.  

\begin{figure}[t]
\begin{center}
\resizebox{8.5cm}{!}{\includegraphics{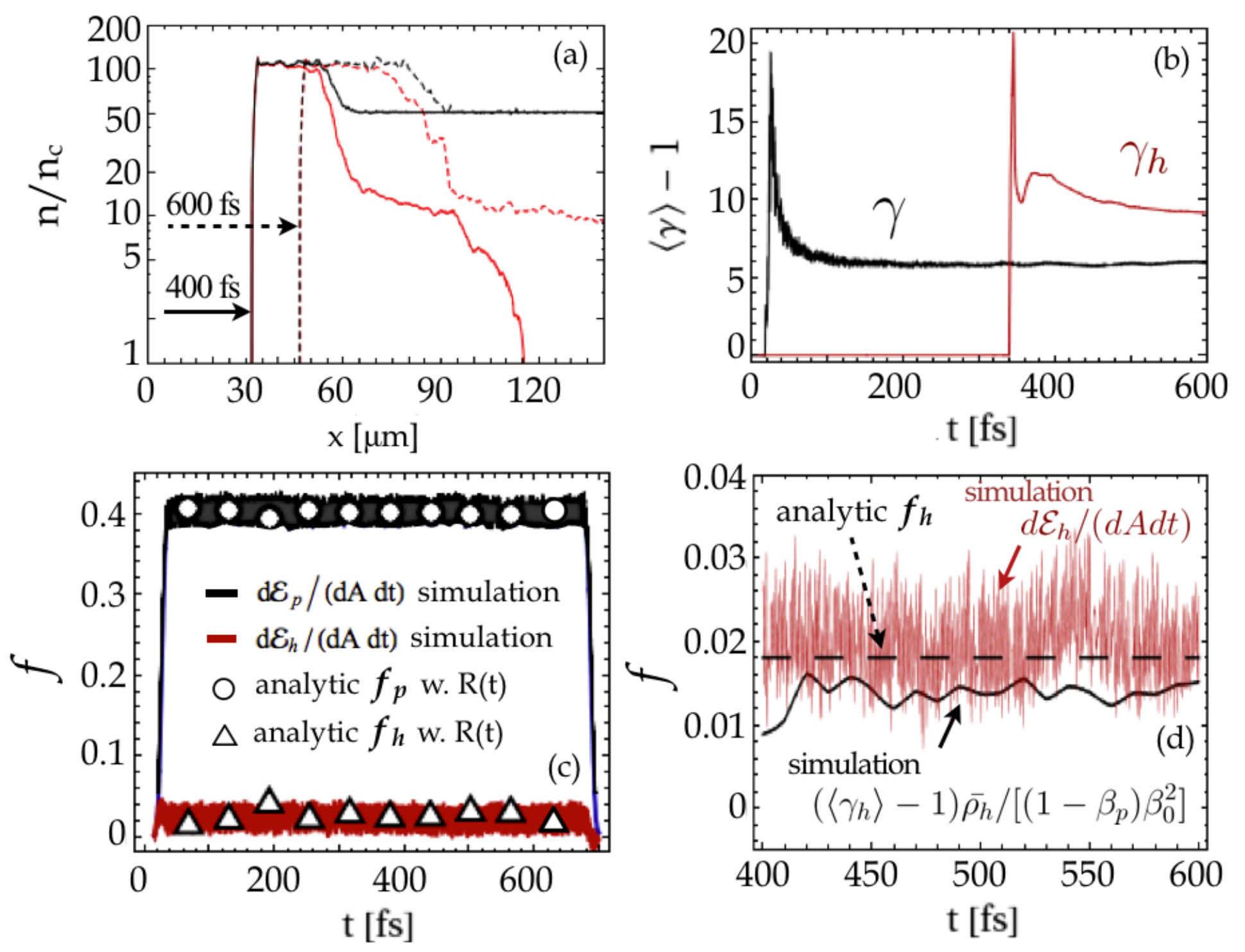}}
  \end{center}
  \caption{Comparison to particle-in-cell simulation results.  (a) Density of exchange-mediating electrons (red) and of ions (black). Arrows indicate the position of the laser-plasma interface.  (b) Ensemble-average Lorentz factors \mcl{for exchange-mediating electrons and hot electrons.} (c) Comparison of laser absorption from the simulation to the analytic model. 
  (d)  Detailed comparison of oscillatory mode coupling.  Simulation parameters are $\bpz=0.35, \dph\ = 0.8$. }
  \label{fig:sims} 
\end{figure}

\mcl{
Finally, we discuss the limitations of our analysis.   The model assumes one dimension in space in steady-state, consistent with scenarios where the laser spot size $r_L$ is large compared to $\lambda_L$, which is readily satisfied in the center of paraxial beams with $r_L \gtrsim $50$\mu$m.  
The assumed damping of transient momentum effects requires that the laser pulse duration $\tau_L$ satisfy the condition $\tau_L \omega_{pi} > 2 \pi A$, where $A\simeq 3-5$.  In addition, the target thickness $x_T$ must exceed the hole punching depth and the effective 'refluxing' hot electron range, $x_T > c\ \tau_L/2 + \int_0^{\tau_L} \bp(t) c\ dt$.}  To maintain steady-state, the laser temporal profile should also change slowly with respect to the electron equilibration timescale, such that $I_L/(\partial I_L/\partial t)^{-1} \omega_{pe} \ll 2\pi$.
}
As the steady-state assumption breaks down, effects such as \mcln{rippling} of the LP interface introduce additional vectors through which the light can couple to the plasma.  These additional degrees of freedom, associated with dynamical and two- and three-dimensional effects, expand the set of exchange-mediating particles in $s$.  In this Letter, our purpose has been to examine the phase aspects of the one-dimensional interaction, thereby establishing bounds on the three-dimensional phenomena. 

\mcl{Equation (\ref{eqn:stressE}) implies that the light interacts uniformly over the relativistic collisionless skin depth as $\ls/\lambda_L \ll 1$.  For hot electrons following the ponderomotive energy scaling\cite{Wilks1992}, the cooling length $\lc$ associated with the radiation reaction force within $\ls$ can be estimated\cite{Esarey1993} as $\lc \simeq 2.1 \times 10^{-2} \sqrt{ \ne [cm^{-3}] }\ a_0^{-7/2} \ \ls$.
Due to the exponential damping of the laser electric field, we calculate that $\lc/\ls \gg 1$ while $I_L \lambda_L^2 < 10^{23}\ W \mu m^2/$cm$^2$ for $\ne/\nc \gtrsim 50$.  Therefore, radiation reaction plays a negligible role.  }

\mcln{
Finally, we comment on multi-dimensional effects in the hot electron velocity distribution.  
$\theta \equiv \tan^{-1} |\mathbf{V_{s2}} \times \hat{k}_L|/ \mathbf{V_{s2}} \cdot \hat{k}_L$ is the half-angle with which hot electrons are \textit{excited} by the laser.  $V_{s2} \rightarrow V_{s2} \cos \theta$ enters into equation (\ref{eqn:stressE}), noting that $\theta$ generally differs from the observed divergence due to downstream effects, e.g., scattering in the plasma bulk, which are abstracted from this analysis. 
 Calculations show that $\fap$ increases with $\theta$, due to the fact that as the projection of the hot electron momentum flux in the axial direction becomes smaller, ions must be pushed faster to conserve momentum with the laser.  Considering the asymptotics, we therefore  see that a strongly-diverging hot electron system must approach the ion-only system described in \cite{Wilks1992,Naumova2009,Macchi2013a}.  This is borne out by the calculations, which exhibit a rapid transition toward $f\rightarrow \fap$ for $\theta \gtrsim 65-75^\circ$.  \mclx{Below this point we find that the absorption limits are qualitatively similar to the nondiverging case, with $\fmax(\theta)\geq \fmax(0)$ and $\falim(\theta)\geq \falim(0)$ at low intensity, converging at high intensity. }For any realistic $\theta$, absorption trends as depicted in Fig. \ref{fig:abs} are therefore maintained.
}

\mclx{
In conclusion, we have derived an inequality bounding the absorption of ultraintense laser light by solids in the regime $10^{18} < I_L \   \lambda_L^2 \lesssim 10^{23}$ W $\mu$m$^2/$cm$^{2}$.  These bounds emerge naturally from constrained optimization techniques applied to a kinematic generalization of the laser-solid interaction as a strongly-driven, relativistic, two degree of freedom Maxwell-Vlasov system subject to boundary conditions. We demonstrate that the extrema and the phase-space-averaged absorption must always increase with intensity, and increase most rapidly when $10^{18} < I_L \   \lambda_L^2 < 10^{20}$ W $\mu$m$^2/$cm$^{2}$.  Our results indicate that the fundamental empirical trend towards increasing fractional absorption with irradiance reflects the underlying phase space constraints.
}

\mcl{M. L. is grateful to B. Breizman} and A. Link for useful discussions.
M. L. acknowledges the LLNL Lawrence Scholarship for support. Computing support for this work came from the LLNL Institutional Computing Grand Challenge program.
This work was performed under the auspices of the U.S. Department of Energy by Lawrence Livermore National Laboratory under Contract DE-AC52-07NA27344.

\end{document}